\documentclass[10pt]{amsart}
\usepackage{enumerate,amsmath,amssymb,latexsym,
amsfonts, amsthm, amscd}

\usepackage{enumerate,amsmath,amsthm,amssymb,latexsym,
amsfonts,amscd}
\usepackage[all]{xy}

\title[The Structure of Local theories of gravity]{The Structure of Local theories of gravity}

\author[Maurice J. Dupr\'e]{M\lowercase{aurice} J. D\lowercase{upr\'e}\\D\lowercase{epartment of} M\lowercase{athematics}
\\N\lowercase{ew} O\lowercase{rleans}, LA 70118\\\lowercase{email:  mdupre@tulane.edu}\\18 S\lowercase{eptember} 2013}
\address{DEPARTMENT OF MATHEMATICS\\TULANE UNIVERSTIY\\NEW ORLEANS, LA 70118}
\email{mdupre@tulane.edu}

\theoremstyle{plain}

\newtheorem{proposition}{Proposition}[section]
\newtheorem{theorem}{Theorem}[section]
\newtheorem{corollary}{Corollary}[section]
\newtheorem{postulate}{Postulate}[section]

\theoremstyle{definition}

\numberwithin{equation}{section}

\newcommand{\B}{\mathcal B}
\newcommand{\Ce}{\mathcal C}
\newcommand{\D}{\mathcal D}

\newcommand{\F}{\mathcal F}

\newcommand{\J}{\mathcal J}
\newcommand{\K}{\mathcal K}

\newcommand{\M}{\mathcal M}

\newcommand{\R}{\mathcal R}
\newcommand{\T}{\mathcal T}
\newcommand{\U}{\mathcal U}

\newcommand{\W}{\mathcal W}

\newcommand{\bC}{\mathbb{C}}

\newcommand{\bK}{\mathbb{K}}
\newcommand{\bN}{\mathbb{N}}

\newcommand{\bR}{\mathbb{R}}

\newcommand{\ra}{\rightarrow}

\newcommand{\lra}{\longrightarrow}

\newcommand{\del}{\partial}

\newcommand{\med}{\medbreak}

\begin{document}

\maketitle

\begin{abstract}
We discuss the structure of local gravity theories as resulting from the idea that locally gravity must be physically characterized by tidal acceleration, and show how this relates to both Newtonian gravity and Einstein's general relativity. 

\end{abstract}

\section{\bf INTRODUCTION}

As a mathematician and outsider to the world of physics, I feel that gives me a perspective which to some extent is above the fray which is taking place in theoretical physics.  Experience shows that successful physical theories follow a fairly well defined sequence of major steps.  The first step is the elucidation of the essential physical ideas in purely physical terms which one needs to describe the theory.  The second step is to describe mathematically a system which corresponds with the physical ideas resulting from the first step.  The third step is to use the mathematical system resulting in step two to make calculations which make predictions of physically interesting quantities.  The fourth step is to experimentally test the predictions resulting from the third step.  One might introduce a fifth step which is to modify physical ideas in step one and go through the sequence of steps again in order to make improvements in the theory.  In short, the steps which can be considered as a cycle, are

\medskip

{\bf (1) Physics

\smallskip

(2) Mathematics

\smallskip

(3) Calculation

\smallskip

(4)  Experiment.}

\medskip

Some comments on these steps are now in order. It is important that the physical ideas of (1) not be overly influenced by mathematics.  In step one, physics must be the main consideration.  In (2) on the other hand, the level of rigour should not be too high as it can prevent progress and get in the way of progress to step three.  Likewise, in (3), for calculations, similarly sometimes physical ideas can be used to aid in calculation where it would not be permitted in pure mathematics.  For instance, if the Weierstrass level of rigour had been required of Newton, it could have prevented the development of Newtonian mechanics.  The reason that the requirements of rigour can be relaxed here is due to the final step (4) which will be the final arbiter of success.  Notice that this means that if (1), (3), and (4) are omitted, all that remains is sloppy mathematics.

With these remarks in mind, let us consider theories of gravity.  Thus, we can say that a theory of gravity is either local or it is not.  As we wish to characterize local theories of gravity, for step one, we will take as our basic physical concept that locally gravity is tidal acceleration.  This means the physical theory we develop merely needs to characterize tidal acceleration, no more.  For step two then, we need a mathematical system $X$ which permits the idea of locality to be expressed and as well, it needs to have the ability to express the kinematics of motion.  Within this system, we need to be able to define a mathematical object which characterizes tidal acceleration, and an object which characterizes whatever is thought to be the source of tidal acceleration.  The theory should then allow the determination of the tidal acceleration object from the source object, for instance via an equation of some sort.  Then, the results of step two should allow the calculations for step three of results for some examples which can be tested experimentally in step four.  With this in mind, we now consider the form of some well known gravity theories.

\section{\bf NEWTONIAN GRAVITY}

Even though gravity has not always been thought of as a local concept, in fact it really has been from the beginning.  It was not until Newton's realization that gravity is responsible for celestial motions that it was viewed globally.  Galileo's experiment in the hold of a ship is really the verification that locally gravity has a universal effect on matter.  What effect gravity does have locally is only in the form of tidal acceleration.  If Galileo had been able to make measurements with enough accuracy, the tidal acceleration effects would have been discovered in the hold of the ship or in dropping balls from towers.  However, Newton's inverse square law of gravity is not a local law, but rather a law of global form.  It is the Poisson equation
\begin{equation}\label{Poisson1}
\Delta V=4 \pi G \rho
\end{equation}
which is equivalent to Newton's theory of gravity which shows that Newtonian gravity is actually a local theory.  Here, of course, $V$ is the Newtonian gravitational potential, and $\Delta$ denotes the Laplace operator on three dimensional Euclidean space, $E,$ and $\rho$ is the matter density.  The functions $V$ and $\rho$ are functions on $E  \times  \bR.$  In Newton's theory of gravity, we really seek to arrive at the gravitational acceleration field ${\bf f}$ which is a function from $E \times \bR$ to $E.$  The potential function $V$ is related to ${\bf f}$ by

\begin{equation}\label{Nacc}
{\bf f}=-\mbox{grad }V, ~\mbox{ at constant time,}
\end{equation}
and as pointed out in \cite{MTW}, the tidal accelerations are completely described by the directional derivatives of the function ${\bf f}.$  That is, $\nabla {\bf f}$ completely determines all tidal accelerations.  Here $\nabla$ is the gradient operator on $E,$ so time is held constant when it operates on functions defined on $E \times \bR.$  We will therefore henceforth call $-\nabla {\bf f}$ the tidal acceleration operator of Newtonian gravity. Thus, in terms of our general considerations, (\ref{Poisson1}) should really be expressed as

\begin{equation}\label{Poisson2}
-\mbox{div } {\bf f}=4 \pi G \rho, ~~{\bf f}=- \mbox{grad } V.
\end{equation}
However, for functions on $E \times \bR,$ the condition that ${\bf f}=-\nabla V$ can be replaced by the symmetry condition that $\nabla_u {\bf f} \cdot v=\nabla_v {\bf f} \cdot u,$ for all $u,v$ in $E,$ which is clearly equivalent to the condition that curl ${\bf f} =0.$  If $L(E;E)$ denotes the vector space of linear 
transformations of $E$ to itself, then $\nabla {\bf f}$ can be viewed as an $L(E;E)$ valued function on $E \times \bR$ whose values are symmetric, that is, self-adjoint.  Mathematically, $\nabla {\bf f}$ is none other than the (partial) Frechet 
derivative of ${\bf f},$ holding time constant.  The special case of the Hodge Theorem that any vector field vanishing at infinity on $E$ is determined by knowing both its divergence and its curl then suffices to determine ${\bf f}$ from $\rho$ via (\ref{Poisson2}).  This in turn then determines $\nabla {\bf f}.$  Notice that we can finally recast the gravitation equation in terms of the tidal acceleration operator ${\bf A}=-\nabla {\bf f}$ as

\begin{equation}\label{Poisson3}
\mbox{trace }{\bf A}=4 \pi G \rho, ~{\bf A}:E \times \bR \lra L(E;E) \mbox{ is symmetric (self-adjoint) valued.}
\end{equation}
Notice that in this final formulation, the equation of Newtonian gravity is simply an equation for determining the tidal acceleration operator which is what we wanted to do in step one purely on physical considerations.  We can also notice that in the view of Newtonians that the object to be determined is the actual acceleration field ${\bf f},$ that in fact as ${\bf A}=-\nabla {\bf f},$ any constant can be added to ${\bf f}$ and the same tidal acceleration operator results.  That is, the essence of tidal acceleration is the variation of the acceleration field from point to point.    This indicates that a uniform acceleration over all of $E$ does not really represent any "real gravity".  Likewise, a constant can be added to the gravitational potential $V$ and the same acceleration field results.   The full tidal acceleration operator ${\bf A}$ is determined by the gravitation equation together with the boundary condition that it vanish at infinity through the Hodge theorem.  From the view that matter is the source of gravity, one should view the equation of gravity as having the matter on the right hand side.  The tidal acceleration operator is merely a kinematic mathematical object.  Thus, I have avoided calling $V$ the gravitational potential energy, as the energy in (\ref{Poisson1}) is in the mass density on the right hand side, not the left hand side.  The left side is merely kinematic, as we see in (\ref{Poisson3}).

\section{\bf  GENERAL RELATIVITY}

Again, we take as our physical idea for step one, that the object of the theory should be to determine the tidal accelerations.  In general relativity (GR), the space $E \times \bR$ of Newtonian dynamics is replaced by spacetime, a general Lorentz manifold $M$ with metric tensor ${\bf g}$ of signature $(-+++).$  Any kinematic allowable motion is a curve in $M$ with velocity vector field $u$ a unit timelike vector field, ${\bf g}(u,u)=-1,$ along the curve.  The acceleration is then the covariant derivative $\nabla_u u$ along the curve of motion.  Here, $\nabla$ denotes the covariant derivative operator for the Levi-Civita connection on $M$ determined by the metric tensor.   The motions physically allowed must obey the law of motion

\begin{equation}\label{lawofmotion}
\nabla_u p=F, ~p=\mbox{ momentum},~F=\mbox{ force}, ~u=\mbox{ unit timelike velocity}.
\end{equation}
The momentum of a particle is $p=mu$ where $m$ is the particle's mass and $u$ is the particle's unit timelike velocity vector.  Thus, a particle free of forces follows a geodesic, $\nabla_u u=0.$

To characterize tidal acceleration here, we have to deal with spacetime curvature.  The Riemann curvature operator, ${\bf R},$ is the  linear transformation valued tensor given on tangent vector fields by
\begin{equation}\label{Riemann1}
{\bf R}(v,w)=[\nabla_v,\nabla_w]-\nabla_{[v,w]},  ~v,w \mbox{ any local smooth tangent vector fields.}
\end{equation}
It is easy to see how to characterize tidal acceleration in this setting.  Given a point $q$ in $M,$ we consider a two-dimensional surface covered by geodesics passing through $q$ and through points near $q,$ resulting in a pair of vector fields $u$ and $s$ defined in an open neighborhood of $q$ with $[u,v]=0,$ and with $u$ being the velocity vector field of the geodesics and $s$ the separation vector, given by the rate of change of position transverse to the geodesics.  The integral curves of $u$ are  thus geodesics, so $\nabla_u u=0.$  The tidal velocity of separation of nearby geodesics is given by the rate of change of separation, $\nabla_u s,$ so the tidal acceleration of nearby geodesics is $\nabla_u^2 s.$  Since $\nabla$ is the covariant differentiation operator for the Levi-Civita connection, it follows that

$$[v,w]=\nabla_v w -\nabla_w v, \mbox{ for any vector fields } v,w,$$
so
$$\nabla_u s=\nabla_s u.$$
Thus,

$${\bf R}(u,s)u=\nabla_u \nabla_s u=\nabla_u^2 s,$$
and this gives the equation of geodesic deviation

\begin{equation}\label{gd1}
{\bf R}(u,s)u=\nabla_u^2 s, \mbox{ or } {\bf R}(s,u)u=-\nabla_u^2 s
\end{equation}
This means that the tidal acceleration operator is $-\nabla_u^2$ for geodesics at $q$ in the direction $u,$ and it is related immediately to the Riemann curvature operator.  But, we need to have the tidal acceleration operator expressed as a function of $u(q)$ which operates on the separation vector $s(q).$  To do this, we form the operator valued tensor ${\bf K}$ equivalent to the Riemann curvature operator defined by

\begin{equation}\label{J1}
[{\bf K}(v,w)]x=[{\bf R}(x,v)]w, \mbox{ for any tangent vectors }v,w,x \mbox{ at the same point.}
\end{equation}
In terms of ${\bf K},$ the equation of geodesic deviation becomes

\begin{equation}\label{gd2}
[{\bf K}(u,u)]s=-\nabla_u^2 s,
\end{equation}
and this means that ${\bf K}$ contains all the tidal acceleration information, and is completely equivalent to the Riemann curvature operator.  However, identities satisfied by the Riemann curvature tensor in fact guarantee that ${\bf K}$ satisfies the identity

\begin{equation}\label{adjoint1}
{\bf K}(x,y)^*={\bf K}(y,x), \mbox{ for any tangent vectors } x,y \mbox{ at the same point of }M.
\end{equation}
Moreover, as only ${\bf K}(u,u)$ enters into the equation of geodesic deviation, (\ref{gd2}), we can in fact replace ${\bf K}$ by its symmetrization which is in fact the {\bf Jacobi curvature operator}, denoted by ${\bf J}.$  Thus,

\begin{equation}\label{J2}
{\bf J}(v,w)=\frac{1}{2}[{\bf K}(v,w)+{\bf K}(w,v)]
\end{equation}
Our equation of geodesic deviation in this final form is simply

\begin{equation}\label{gd3}
{\bf J}(u,u)s=-\nabla_u^2 s
\end{equation}

Thus, finally, the Jacobi curvature operator is the true tidal acceleration operator in GR, it is a symmetric linear transformation valued tensor whose values are all symmetric  ( in other words, self-adjoint) linear transformations, as a result of (\ref{adjoint1}).  In particular, we see that in GR, each observer must have his own tidal acceleration operator.  The various identities satisfied by the Riemann curvature operator in fact also guarantee that it is completely determined by the Jacobi curvature operator.  Thus each determines the other.  In a sense, the curvature can either be determined by a completely anti-symmetric object, the Riemann curvature operator, or a completely symmetric object, the Jacobi curvature operator.  From the viewpoint of gravity as a local theory being determined by tidal acceleration, we see that for step two in GR, we need to deterimine the Jacobi curvature operator.

Here, the various identities such as Bianchi identities and symmetries, together with boundary conditions specified on a Cauchy surface will guarantee that it suffices to determine the trace of the tidal acceleration operator.  The trace of the tidal acceleration operator is in fact simply the Ricci curvature tensor, ${\bf Ricci},$ that is,

\begin{equation}\label{Ricci1}
{\bf Ricci}(v,w)=\mbox{ trace }[{\bf J}(v,w)], \mbox{ any tangent vectors } v,w \mbox{ at a point.}
\end{equation}

Now, going back to step one for the moment, physically, the tidal acceleration operator should be determined in GR, by all forms of energy, so as ${\bf Ricci}$ is a second rank symmetric tensor, the equation along the lines of Newtonian gravity should simply be

$${\bf Ricci}=4 \pi G ~{\bf S},$$
where ${\bf S}$ is a second rank symmetric tensor which characterizes all the forms of energy that influence gravity.  From this standpoint, we can now see what the Einstein equation says.  It says that

\begin{equation}\label{Einstein1}
{\bf Ricci}=4 \pi G [{\bf T}+({\bf T}-T{\bf g})]
\end{equation}
where ${\bf T}$ is the energy momentum stress tensor of all matter and fields other than gravity, and where $T$ is the contraction of ${\bf T}.$  We notice here that the matter tensor enters in a second term ${\bf S}_g,$ where

\begin{equation}\label{gravnrg1}
{\bf S}_g={\bf T}-T{\bf g}.
\end{equation}
What do we make of this tensor ${\bf S}_g$?  Since ${\bf T}$ is an energy momentum stress tensor, in terms of units alone, we must have that ${\bf S}_g$ is also an energy momentum stress tensor of something.  Since it is symmetric, it is completely determined by its monomial form on timelike unit vectors.  That is, ${\bf S}_g$ is completely determined by its energy ${\bf S}_g(u,u)$ as seen by each observer with unit timelike velocity $u,$ as noted in \cite{SACHSWU} and more generally in \cite{DUPRE} and \cite{DUPRE3}.  An easy calculation now shows that for any timelike unit vector $u,$ the quantity ${\bf S}_g(u,u)$ is the sum of the principal pressures as seen by an observer with velocity $u.$  This means that pressures enter into the source of gravity as energy in addition to that already present in ${\bf T}$ itself.  At this point, it is natural to consider that the only thing that could be giving this extra energy is the gravitational field itself, so ${\bf S}_g$ should be considered as the energy momentum stress tensor of the gravitational field itself.  Consequently,

\begin{equation}\label{nrg1}
\frac{1}{4 \pi G} {\bf Ricci}=\mbox{\bf Energy Momentum Stress Tensor of all Matter and Fields Including Gravity}.
\end{equation}
We can also see that arguments which give ${\bf S}_g$ as the energy momentum stress tensor of gravity in turn give a direct argument for the Einstein equation (\ref{Einstein1}), which in turn implies that the divergence of ${\bf T}$ must be zero, a fact which is often assumed at the outset in derivations of the Einstein equation, (\ref{Einstein1}).  For instance in \cite{DUPRE} and \cite{DUPRE2}, simple arguments are given for the fact that the sum of principal pressures should be viewed as the energy density of the gravitational field, as seen by any observer.  From the standpoint of viewing ${\bf S}_g$ as the energy of the aether, see \cite{D&T1}.  From the standpoint of quasi-local energy, see \cite{C&D2} for the use of ${\bf Ricci}$ to obtain energy of extended systems, called spacetime energy momentum.

\section{\bf OTHER SPACETIME GRAVITY THEORIES}

There are at this point, too many theories of gravity for spacetime to possibly reference them all.  However, they almost all have in common the fact that the Ricci tensor is part of the equation.  Since the Ricci tensor and boundary conditions determine the metric tensor and the Jacobi curvature operator which is the tidal acceleration operator, in effect, these theories are seen to be introducing new sources of gravity.  In higher dimensional gravity theory, it might take more than the trace of the Jacobi curvature operator and boundary conditions to determine the full Jacobi curvature operator, so that other invariants other than the trace of the Jacobi curvature operator could conceivably enter the theory on the geometric side of the equation.  However, for ordinary spacetime of signature $(-+++),$ only the trace of the Jacobi curvature operator need enter the equation, as the rest is determined by the boundary conditions.  Consequently, any thing in a spacetime gravity equation other than the Ricci tensor should be regarded as part of the source of gravity in such a theory and put on the source side of the equation.

\section{\bf ACKNOWLEDGEMENTS}

Even though they cannot rightfully be blamed for any ideas expressed here, I must thank and acknowledge Fred Cooperstock and Frank Tipler for many useful and informative conversations and emails concerning the developments contained here.

\end{document}